\documentclass{article}[12pt]

\usepackage{epsfig}

\begin{document}

\centerline{\huge Evolving eco-system: a network of networks}

\bigskip
Debashish Chowdhury$^1$ and Dietrich Stauffer$^2$

\bigskip
Department of Physics, Indian Institute of Technology, 
Kanpur 208016, India

\bigskip 
Institut for Theoretical Physics, Cologne University, 
D-50923 K\"oln, Euroland

\bigskip
E-mail: debch@iitk.ac.in, stauffer@thp.uni-koeln.de

\begin{abstract}
Ecology and evolution are inseparable. Motivated by some recent 
experiments, we have developed models of evolutionary ecology from
the perspective of dynamic networks. In these models, in addition to
the intra-node dynamics, which corresponds to an individual-based 
population dynamics of species, the entire network itself changes 
slowly with time to capture evolutionary processes. After a brief 
summary of our recent published works on these network models of 
eco-systems, we extend the most recent version of the model 
incorporating predators that wander into neighbouring spatial 
patches for food.
\end{abstract}

\vspace{0.7cm}


\section{Introduction}

An eco-system may be viewed as a functional network of species; the 
food web \cite{foodweb} corresponding to the eco-system consists of 
nodes and links where each node corresponds to a species and the 
(directed) links represent the prey-predator interactions such that 
the direction of the link indicates the direction of flow of nutrient 
(i.e., {\it from} a prey {\it to} one of its predators) 
\cite{drossel,newman}. For convenience, most of the earlier ecological 
models that describe population dynamics, usually ignored 
macro-evolutionary changes in the eco-system and, therefore, assumed 
the food web to be independent of time. On the other hand, most of the 
macro-evolutionary models \cite{bak,solerev} of speciation and extinction 
of species did not explicitly explore the ageing and age-distributions 
of the populations of various species in the system. The models of ageing, 
usually, focus on only one single species. 

But, recent experimental evidences \cite{thompson99,stockwell03,yoshida03} 
have established that significant evolutionary
changes can occur over ecologically relevant time scales. Motivated 
by these experiments, we have developed models 
\cite{csk,cspre,scpa,cspa,stauetal04} of evolutionary ecology from 
the perspective of dynamic networks. In these models, in addition to 
the intra-node dynamics, which corresponds to population dynamics of 
species, the entire network itself changes slowly with time to capture 
evolutionary processes. The aim of these models is to provide insight 
into the mechanisms that give rise to the generic features of the 
biological evolution of real eco-systems. 

In this paper, after a brief review of the earlier network models, 
including our own \cite{csk,cspre,scpa,cspa}, we extend the most 
recent version \cite{stauetal04} of our model by allowing predators 
to prey on species in the neighbouring spatial patches as well.

\section{Earlier network models} 

A network model of ecosystems was developed by Sole and Manrubia
\cite{sole}. The state of the $i$-th species ($i = 1,2,...N$) is 
represented by a two-state variable $S_i$; $S_i = 0$ or $1$ 
depending on whether it is extinct or alive, respectively. The 
inter-species interactions are captured by the interaction matrix 
${\bf J}$; the element $J_{ij}$ denotes the influence {\it of} 
the species $j$ {\it on} the species $i$.  If $J_{ij} > 0$ while, 
simultaneously, $J_{ji} < 0$ then $i$ is the predator and $j$ is 
the prey. On the other hand, if both $J_{ij}$ and $J_{ji}$ are 
positive (negative) they the two species cooperate (compete). The 
food web in the Sole-Manrubia model \cite{sole} has a random 
architecture.

The dynamics of the system consists in updating the states of the
system in three steps. At the first step, one of the input 
connections $J_{ij}$ for each species $i$ is picked up randomly 
and assigned a new value drawn from the uniform distribution in 
the interval $[-1,1]$, irrespective of its previous magnitude and 
sign. At the second step, the new state of each of the species is 
decided by the equation
\begin{equation}
S_i(t+1) = \Theta \biggl(\sum_{j=1}^N J_{ij} S_j(t) - \theta_i\biggr)
\end{equation}
where $\theta_i$ is a threshold parameter for the species $i$ and
$\Theta(x)$ is the standard step function. If $S(t+1)$ becomes zero 
for $m$ species, then an extinction of size $m$ is said to have taken
place. Finally, at the third step, all the niches left vacant by the 
extinct species are refilled by copies of one of the randomly selected
non-extinct species. Sole and Manrubia \cite{sole} observed that the 
distributions of the sizes of these extinctions could be fitted to a 
power law of the form $N(m) \sim m^{-\alpha}$ with an exponent 
$\alpha \simeq 2.3$. 

Abramson \cite{abramson} considered a linear food web which is 
extremely unrealistic and required a constant number of species.
Amaral and Meyer \cite{amaral} considered a hierarchical food web
but the population dynamics was oversimplified. The main limitation 
of these network models is that the individual organisms do not 
appear explicitly.

\section{The ``unified'' network model}

\begin{figure}[tb]
\begin{center}
\includegraphics[angle=-90,width=0.9\columnwidth]{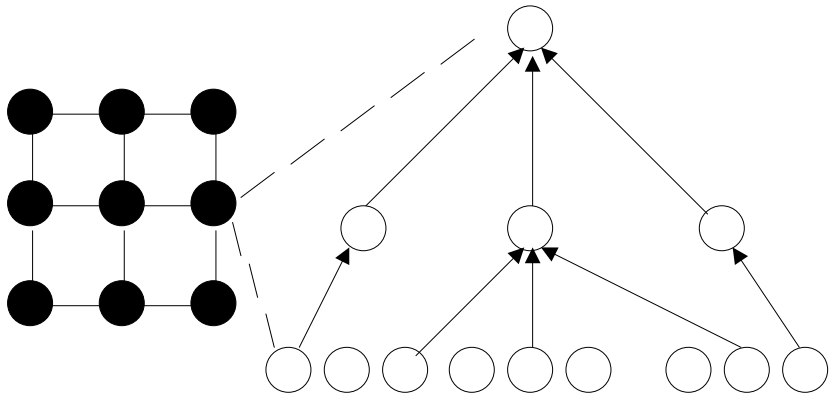}
\end{center}
\caption{A schematic representation of the eco-system by a network 
of networks. Each filled circle in the left part, representing a 
spatial ``patch'' of the eco-system, is endowed with a hierarchical 
network, shown in the right part, that graphically represents the 
food web structure. 
}
\label{fig-netofnet}
\end{figure}

We represent the spatial extensions of the eco-system by a square 
lattice where each site represents a spatial ``patch'' (see the 
left side of Fig.\ref{fig-netofnet}). Moreover, a food web is assigned 
to each spatial ``patch'' (see the right side of Fig.\ref{fig-netofnet}). 
Thus, the eco-system can be modelled as a {\it network of networks} 
(see Fig.\ref{fig-netofnet}).

As in our earlier papers \cite{cspre,scpa,cspa,stauetal04}, we assume 
a generic {\it hierarchical} architecture of the food web, where niches 
are arranged in different trophic levels $\ell$ 
($1 \le \ell \le \ell_{max}$), with no more than $m^{\ell - 1}$
nodes in each level ($m$ is a positive integer). Species in level 
$\ell$ can prey on (or ignore) species on the immediately lower 
level $\ell +1$. The total number of species cannot exceed 
$N_{max} = (m^{{\ell}_{max}}-1)/(m-1)$, the total number of nodes.  
Note that $\ell_{max}$ and $m$ (and, therefore, $N_{max}$) are 
time-independent parameters in the model. At any arbitrary instant 
of time $t$ the model consists of $N(t)$ {\it species} each of which 
occupies one of the nodes of the dynamic network. Our model allows 
$N(t)$ to fluctuate with time over the range 
${\ell} \leq N(t) \leq N_{max}$.

Following the spirit of the Sole-Manrubia model \cite{sole} 
we describe the prey-predator interaction by the elements of the 
interaction matrix $J$, except that, for simplicity, we allow only 
the discrete integer values $J_{ij} = \pm 1, 0$.  

The food web itself evolves slowly over sufficiently long time scales. 
In order to capture the changing food habits, each of the species in 
our model re-adjusts, with the probability $p_{mut}$, a link $J$ from 
one of its predators and another to one of its potential preys at 
every time step \cite{sole}. Moreover, random genetic mutations are 
captured also by implementing random tinkering of some of the 
intra-node characteristics which will be introduced below. 
Furthermore, even the occupants of the nodes can change 
with time because, following extinction, the vacant nodes are 
slowly re-occupied through speciation, to be explained below.

Population dynamics at the neighbouring patches are coupled by {\it 
migration}: a population may expand into a neighbouring lattice site, 
if the population there is zero for the same species on the same level. 
A direct inter-``patch'' interaction enters through the wandering of 
the predators into neighbouring patches for food: a species at a patch 
$i,j$ can prey on species at the next lower level in the food web 
located the neighbouring patches $i \pm 1,j$ and $i, j \pm 1$ as well 
as those at the same location, namely, $i,j$.

The {\it intra}-species competitions among the organisms of the same 
species for limited availability of resources, other than food, 
imposes the upper limit $n_{max}$ of the allowed population of each 
species. The population (i.e., the total number of organisms) of a 
given species, say, $i$, in the spatial patch $\alpha$ at any arbitrary 
instant of time $t$ is given by $n_{i,\alpha}(t) \leq n_{max}$. 
Thus, the total number of organisms $n(t)$ in the eco-system at time 
$t$ is given by $n(t) = \sum_{i=1}^{N(t)} \sum_{\alpha} n_{i,\alpha}(t)$. 

For simplicity, we assume the reproductions to be {\it asexual}.
At each time step, the survivors give 
birth to $M$ offspring with probability
$$p_{birth} = [(X_{max}-a)/(X_{max}-X_{rep})][1- n/n_{max}]$$
if their age $a$ is above the minimum reproduction age $X_{rep}$.

We assume that each individual either ages by one time unit for
each time step, or it dies. In addition to the possibility of 
death as prey, the probability of natural death of each organism  
of age $a$ is assumed to be given by 
$$p_{death} = \exp[(a-X_{max})r/M]$$
where $X_{max}$ is the maximum possible age and $M$ is the litter 
size of the whole species; where $r$ is a free parameter, e.g. 0.05. 
(For ages below the minimum reproduction age $X_{rep}$ the death 
probability is assumed to be age-independent, with $a$ replaced
by $X_{rep}$ in the above equation.)

During each time step, because of random genetic mutations, $X_{rep}$ 
and $M$ independently increase or decrease by unity, with equal probability, 
$p_{mut}$.  $X_{rep}$ is not allowed to exceed a $X_{max}$ of this 
species, while $M$ is restricted to remain positive.

The $J$ account not only for the {\it inter}-species interactions but 
also {\it intra}-species competitions for food. Let $S_i^+$ be the 
number of all prey individuals for species $i$ on the lower trophic 
level, and $S_i^-$ be $m$ times the number of all predator individuals 
on the higher trophic level. Because of the larger body size of the 
predators, we assume that a predator eats $m$ prey per 
time interval. Then, $S_i^+$ gives the available food for species $i$, 
and $S_i^-$ is the contribution of species $i$ to the available food 
for all predators on the next higher level. 
If $n_i-S_i^+$ is larger than $S_i^-$ then food shortage will be the 
dominant cause of premature death of a fraction of the existing 
population of the species $i$. On the other hand, if $n_i-S_i^+ < S_i^-$, 
then a fraction of the existing population will be wiped out primarily 
by the predators.

Because of the natural death mentioned above and, more importantly, 
prey-predator interactions, the populations of some species may fall 
to zero. In order to capture the process of {\it speciation}, all the 
empty nodes  in a trophic level of the network are re-filled, 
with a probability $p_{sp}$, by random mutants of {\it one common 
ancestor} which is picked up randomly from among the non-extinct 
species at the same trophic level. The subsequent accumulation of 
random mutations over sufficiently long time leads to the divergence 
of the genomes of the parent and daughter species that is an essential 
feature of speciation.

\begin{figure}[tb]
\begin{center}
\includegraphics[angle=-90,width=0.9\columnwidth]{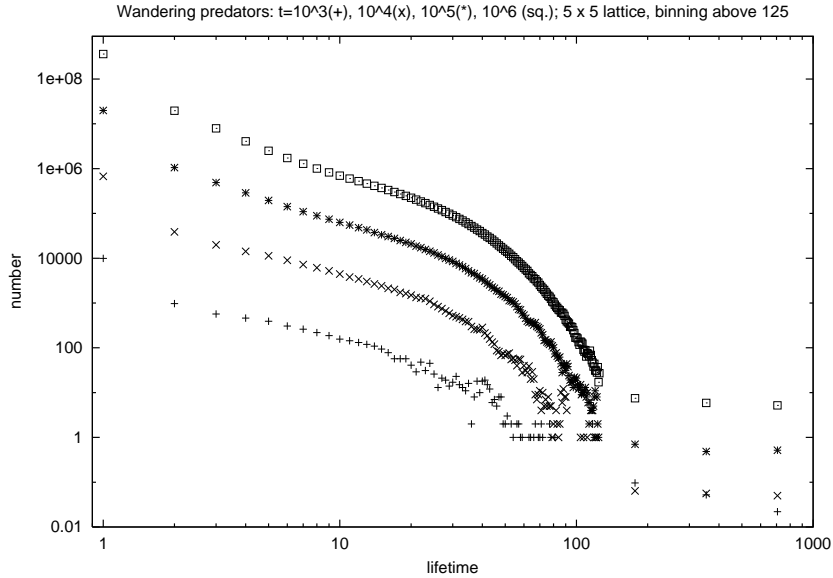}
\end{center}
\caption{Log-log plots of the distributions of the lifetimes of the
species. The common parameters for all the curves are
$m=2$, $p_{lev} = 0.001$, $p_{sp} = 0.1$, $p_{mut} = 0.001$, 
$C = 0.2, r = 0.05$. 
The symbols $ +, \times$, $\ast$ and the unfilled square correspond to
$t = 10^3, 10^4, 10^5, 10^6$ on a $5 \times 5$ square lattice. 
}
\label{fig-2}
\end{figure}

However, occasionally, all the niches at a level may lie vacant. 
Under such circumstances, all these vacant nodes are to be filled 
by a mutant of the non-extinct species occupying the closest {\it 
lower} level. In our computer simulations, the search for this 
non-extinct species is carried out in steps, if even the lower 
level is also completely empty, the search for survivor shifts to 
the next lower level and the process continues till the lowest 
level is reached. The species at the lowest level (representing, 
for example, plankton) are assumed to be immortal.

In order to capture the fact that real ecosystems can exhibit 
growing bio-diversity over sufficiently long period of time, we 
allowed adding a new trophic level to the food web, with a small 
probability $p_{lev}$ per unit time, provided the total bio-mass 
distributed over all the levels (including the new one) does not 
exceed the total bio-mass available in the eco-system. Increase of 
the number trophic level means the diversification at the erstwhile 
topmost level as well as all the lower levels and the emergence of 
yet another dominating species that occupies the new highest level.

\section{Results and conclusions}

The average distributions of the lifetimes of the species are plotted 
in fig.\ref{fig-2}. It is not possible to fit a straight line through 
the data over the entire range of lifetimes; only a limited 
regime is consistent with a power-law with the effective exponent 
$2$, which has been predicted by several models of ``macro''-evolution 
\cite{drossel,newman}. This qualitative behaviour is similar to those 
observed earlier with simpler versions of our ``unified'' model, 
except for the new feature that a plateau appears in the Fig\ref{fig-2} 
for lifetimes $\gg 100$.

Because of the various known limitations of the available fossil data, 
it is questionable whether real extinctions follow power laws and, if 
so, over how many orders of magnitude.

In summary, we have extended the most recent version of our model ``unified'' 
model of evolutionary ecology \cite{csk,cspre,scpa,cspa,stauetal04}, 
formulated as a network of networks, by incorporating predator-prey 
interactions among species on neighbouring spatial patches.
This improvement does 
not alter the qualitative features of the statistics of extinctions 
in our ``unified'' model.

\bigskip

\noindent {\bf Acknowledgements} 

We thank Ambarish Kunwar for enjoyable collaborations and DFG/BMZ for 
supporting this work through the Indo-German research grant DFG/Sta130.

\end{document}